\documentclass[twocolumn, superscriptaddress, secnumarabic, amssymb, showpacs, nobibnotes, aps, prb]{revtex4-1}
\usepackage{graphicx }
\usepackage{dcolumn}
\usepackage{bm}
\usepackage{xcolor}
\usepackage{upgreek}
\usepackage{xspace}
\usepackage{amsmath}

\begin{document}
\newcommand{\fp}{Fe$_3$PO$_4$O$_3$}
\newcommand{\muSR}{$\mu$SR\xspace}
\newcommand{\mus}{$\upmu$s\textsuperscript{-1}\xspace}
\newcommand{\REF}{{\bf[REF]}}

\title{
Magnetic order and spin dynamics in the helical magnetic system  Fe$_3$PO$_4$O$_3$
}
\author{R. Sarkar}
\email{rajibsarkarsinp@gmail.com}
\author{S. Kamusella}
\author{S. A. Br\"auninger}
\affiliation{Institute of Solid State and Materials Physics, TU-Dresden, 01062 Dresden, Germany}
\author{S. Holenstein}
\affiliation{Laboratory for Muon Spin Spectroscopy, Paul Scherrer Institute, CH-5232 Villigen PSI, Switzerland}
\affiliation{Physik-Institut der Universitat Z\"urich, Winterthurerstrasse 190, CH-8057 Z\"urich, Switzerland}
\author{J.-C. Orain}
\author{H. Luetkens}
\affiliation{Laboratory for Muon Spin Spectroscopy, Paul Scherrer Institute, CH-5232 Villigen PSI, Switzerland}

\author{V. Grinenko}
\affiliation{Institute for Solid State Physics, TU Dresden, D-01069 Dresden, Germany}

\author{M. J. Tarne}
\author{J. R. Neilson}
\affiliation{Department of Chemistry, Colorado State University, Fort Collins, Colorado 80523-1872, USA}

\author{K. A. Ross}
\affiliation{Department of Physics, Colorado State University, Fort Collins, Colorado 80523-1875, USA}

\author{H.-H. Klauss}
\affiliation{Institute for Solid State Physics, TU Dresden, D-01069 Dresden, Germany}

\date{\today}

\begin{abstract}

The 3$d$-electronic spin dynamics and the magnetic order in \fp\ were investigated by muon spin rotation and relaxation ($\mu$SR) and $^{57}$Fe M\"ossbauer spectroscopy.
Zero-field (ZF)-$\mu$SR and $^{57}$Fe M\"ossbauer studies confirm static long range magnetic ordering below $T_{\mathrm{N}}$  $\approx$ 164\,K. Both transverse-field (TF) and ZF-\muSR\ results evidence 100\% magnetic volume fraction in the ordered state. The ZF-\muSR\ time spectra can be best described by a Bessel function, which is consistent with the helical magnetic structure proposed by neutron scattering experiments. The M\"ossbauer spectra are described in detail by considering the specific angular distribution of the local hyperfine field $B_{\mathrm{hyp}}$ with respect to the local electric field gradient. The \muSR\ spin-lattice relaxation rate exhibits two peaks: One at the magnetic ordering temperature related to critical magnetic fluctuations and another peak at 35\,K signaling the presence of a secondary low energy scale in \fp.

\end{abstract}

                          
\maketitle

The study of helical magnets is an important research topic in strongly correlated electron systems, in part due to their ability to form Skyrmions.~\cite{Tokura-2013}  Skyrmions are topological spin textures that may arise from a helical state either in the form of a Skyrmion lattice phase or as individual Skyrmions produced as defects in the helical structure (for instance, by manipulating domain walls).~\cite{{Jonietz1648},{Junichi-2013},{Uchida359}}  Several non-centrosymmetric helimagnets from the B20 structure type, such as MnSi\cite{Muhlbauer915-FeSi}, FeSi~\cite{Muhlbauer915-FeSi}, and Fe$_{1-x}$Co$_{x}$~\cite{PhysRevB.81.041203-FeCoSi} are known to support Skyrmion lattice phases.  The lack of an inversion center in these materials enables a sizable Dzyaloshinskii-Moriya interaction, which in these materials competes with ferromagnetic Heisenberg exchange to form the helical states.

\fp\ is a non-centrosymmetric material (space group $R3m$)~\cite{{Gavoille-1987},{cipriani1997}} in which an unusual incommensurate magnetic structure developes below $T_N$=163 K.~\cite{PhysRevB.92.134419} \fp\ has been previously studied by heat capacity,~\cite{{Shi201386},{PhysRevB.92.134419}} magnetization,~\cite{{Gavoille-1987},{JSSC-1983},{PhysRevB.92.134419}} neutron powder diffraction (NPD)~\cite{{Gavoille-1987},{PhysRevB.92.134419}}, and $^{57}$Fe M\"ossbauer spectroscopy measurements.~\cite{Gavoille-1987} Susceptibility measurements reveal that \fp\ is a frustrated antiferromagnet with a frustration index higher than 6.~\cite{PhysRevB.92.134419} Early reports on \fp\ proposed a commensurate magnetic structure based on NPD, though the broadness of some diffraction features remained puzzling at the time.~\cite{Gavoille-1987}  M\"ossbauer data were also reported down to 77\,K,~\cite{Gavoille-1987} which confirmed static magnetic order below $T_N$.  However, there was insufficient accuracy of the M\"ossbauer data to resolve the magnetic structure of \fp\, or the relevant magnetic order parameter.~\cite{{cipriani1997},{JSSC-1983},{Gavoille-1987}} Recent NPD measurements found an incommensurate modulation of an antiferromagnetic (AFM) parent structure, with the incommensurate wave-vector lying in the hexagonal $ab$ plane.~\cite{PhysRevB.92.134419}  Surprisingly, the correlations in this plane were found to be restricted to $\xi_{ab} \approx$ 70 nm down to the lowest measured temperatures (4\,K), while long range commensurate correlations along $c$ were simultaneously observed. This hints at the presence of a high density of domain walls separating needle-like helical domains, which in turn suggests the possibility of observing topological spin structures such as AFM Skyrmions in this material.~\cite{{Zhang-2016},{PhysRevLett.116.147203-skyrmion},{Zhang-NC-2016}}

To further investigate this unusual material, particularly its spin dynamics and the nature of the static magnetic order, we have performed measurements on \fp\ over a range of timescales using both macroscopic (ac susceptibility) and local probes (\muSR\ and M\"ossbauer spectroscopy). We present detailed M\"ossbauer spectroscopy results in the temperature range 4.2 - 295\,K, and \muSR\ measurements in the temperature range 2.6 - 295\,K.  $^{57}$Fe M\"ossbauer and zero-field (ZF)-$\mu$SR studies confirm the static long range magnetic ordering below $T_{\mathrm{N}}$ $\approx$ 164 $\pm$1\,K. These data are in agreement with the helical magnetic structure suggested by the neutron scattering results. M\"ossbauer, transverse-field (TF) and zero-field (ZF) \muSR\ studies evidence a 100\% magnetic volume fraction deep in the magnetically ordered state. The \muSR\ spin-lattice relaxation rate displays a peak at~163\,K due to critical fluctuations surrounding the magnetic transition. Unexpectedly, a second peak occurs at 35\,K indicating the presence of a secondary low energy scale in \fp. We discuss possible reasons for this low temperature dynamics. Further, on the basis of the M\"ossbauer data, we propose different micro-magnetic models of the magnetic structure to describe the M\"ossbauer data, and verify the limitations or the feasibility of using these micro-magnetic models to describe the magnetic structure of \fp.

\section{Experimental}

A polycrystalline sample of \fp\ was prepared at the Colorado State University by standard solid-state methods as described in Ref [\onlinecite{PhysRevB.92.134419}]. \muSR\ experiments were performed at the Paul Scherrer Institute, Switzerland using the GPS and DOLLY instruments. To optimize the fraction of muons stopping in the 48\,mg of powder sample, a 300 $\mu$m Al foil was used in front of the sample to degrade the kinetic energy of the muons. The \muSR\ data were analyzed with the free software package MUSRFIT.~\cite{SUTER201269}
The ac susceptibility was measured on a Quantum Design, MPMS XL5 SQUID magnetometer. A 121.8 mg specimen of \fp\ was contained within a gelatin capsule held in a polyethylene drinking straw.  The data were collected on cooling in zero applied dc field.  The driving amplitude of the ac field was 0.4\,mT; the measurement was performed at 3 different frequencies (33 Hz, 332 Hz, 999 Hz).

M\"ossbauer measurements were carried out in an Oxford He flow cryostat using a standard WissEl M\"ossbauer spectrometer. We used a 1.4\,GBq Rh/Co-Source and a Si-PIN-detector from KeTek. Spectra were taken at increasing temperatures, measuring not longer than 12 hours for each spectrum. The absorber exceeded the thin absorber limit considerably (effective thickness $t_a\approx 10.6$), requiring a transmission integral analysis. The analysis of the M\"ossbauer spectra was done using Moessfit.~\cite{kamusella2016moessfit}

\section{\muSR\ and ac susceptibility results}

\begin{figure}[h]
\includegraphics[width=\columnwidth]{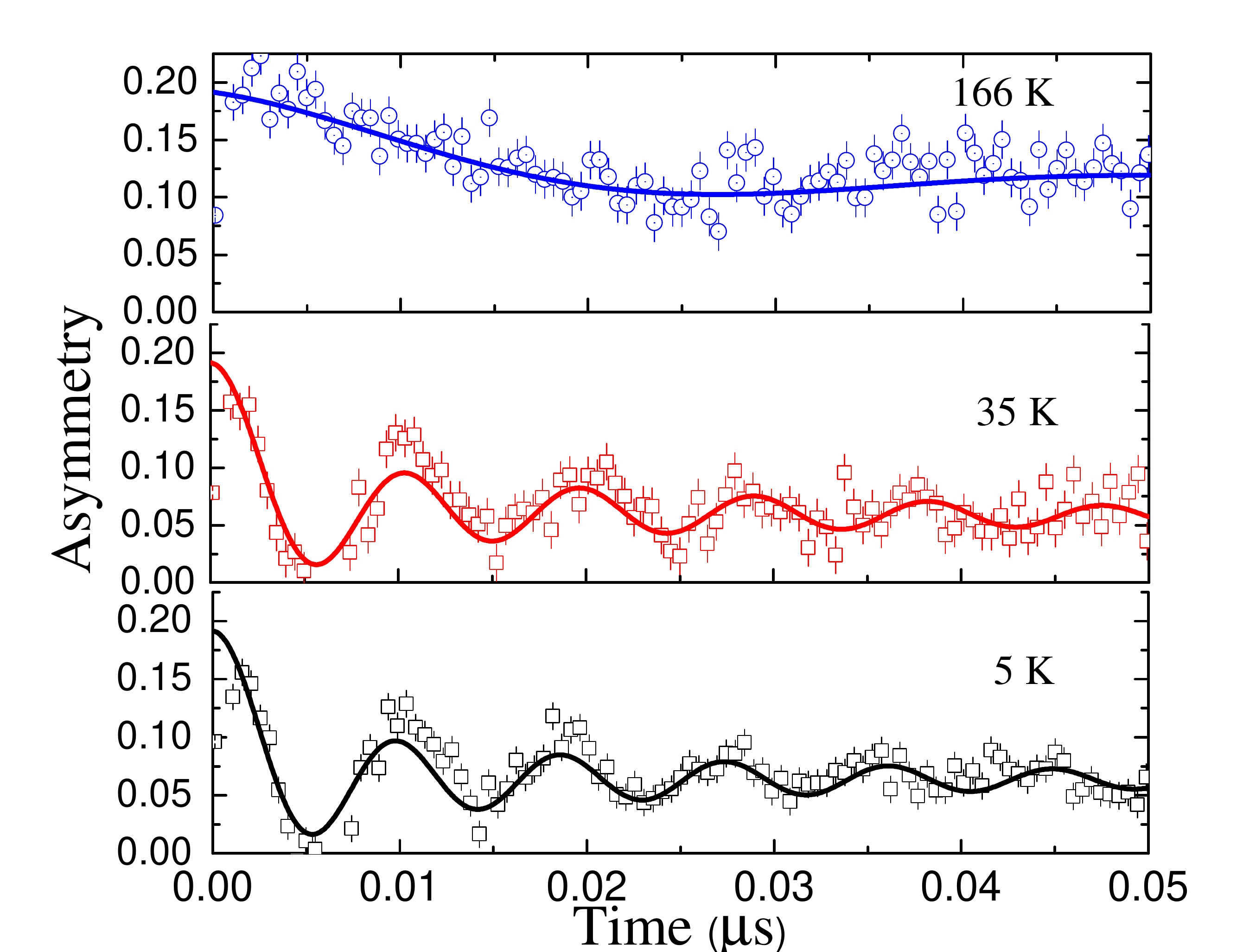}
\caption{\label{fig:early-musr-spectra} ZF-\muSR\ spectra at early decay times below the magnetic ordering temperature $T_{\mathrm{N\mu}}$ = 168\,K measured in GPS. Lines indicate the theoretical description as detailed in the text.}
\end{figure}

\begin{figure}[h]
\includegraphics[width=\columnwidth]{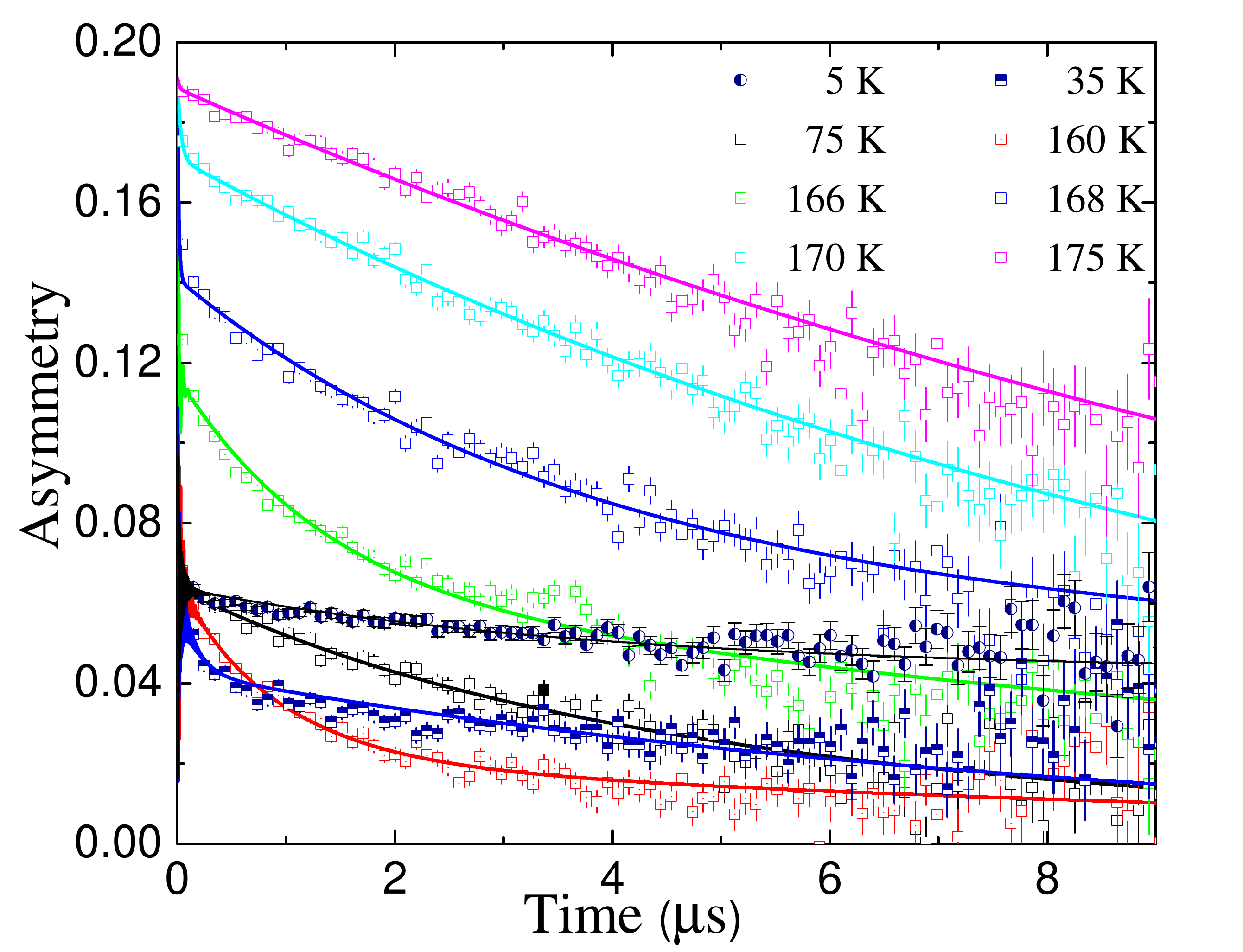}
\caption{\label{fig:late-musr-spectra} ZF-\muSR\ late time spectra both above and below the ordered state measured in GPS. Lines indicate the theoretical description as detailed in the text.}
\end{figure}

Representative ZF-$\mu$SR asymmetry spectra measured in the ordered as well as in the paramagnetic state are shown for short and long times in Fig.~\ref{fig:early-musr-spectra} and \ref{fig:late-musr-spectra} respectively. Spontaneous coherent oscillations below $T_{\mathrm{N\mu}}\approx$ 168\,K (Fig.~\ref{fig:early-musr-spectra}) indicate the presence of a static magnetic field at the muon stopping site. The oscillations can be modelled with a Bessel function, which is consistent with the helical incommensurate structure seen by neutron scattering.~\cite{yaouanc2011muon} To fully describe the spectra, an exponentially damped non-magnetic fraction needs to be included in the transverse part of the asymmetry and the longitudinal part has to be modelled by a sum of two exponential relaxations:

\begin{align}
A(t)=& A_0 \frac{2}{3}[(f_\mathrm{mag}J_{0}(\gamma_\mathrm{\mu}B t)e^{-\lambda_\mathrm{mag}t})
\notag \\
& + (1-f_\mathrm{mag})e^{-\lambda_\mathrm{non-mag}t}]
\notag \\
 & + \frac{1}{3} [f_\mathrm{A}e^{-(\lambda_\mathrm{tail,A}t})+(1-f_\mathrm{A})e^{-(\lambda_\mathrm{tail,B}t})],
 \label{eqn:muonAsymmetry}
\end{align}

where $J_{0}$ represents a Bessel function of the first kind, $B$ is the magnetic field at the muon site (see Appendix \ref{sec:dipole-field} for the possible muon site estimation) and $\gamma_\mathrm{\mu}$ is the gyromagnetic ratio of the muon. The 2/3 (so-called transverse) and 1/3 (so-called longitudinal or tail) terms in equation Eq.~(\ref{eqn:muonAsymmetry}) originate from the powder nature of the sample, i.e. the random orientation of the individual grains w.r.t. the initial muon spin polarization. Given that the helical magnetic domains are randomly oriented and therefore the corresponding magnetic field at the muon site as well, on average, 2/3 of the implanted muons precess around a field perpendicular to their spin while 1/3 of the implanted muons experience longitudinal fields and don't precess. Strictly, one would have to apply a two sites model (sites A and B) for the transverse (2/3) part in Eq.~(\ref{eqn:muonAsymmetry}), too, to make it consistent with the applied model for the longitudinal part. But such a model yields identical frequencies and similar relaxation rates for both sites.

\begin{figure}[h]
\includegraphics[width=\columnwidth]{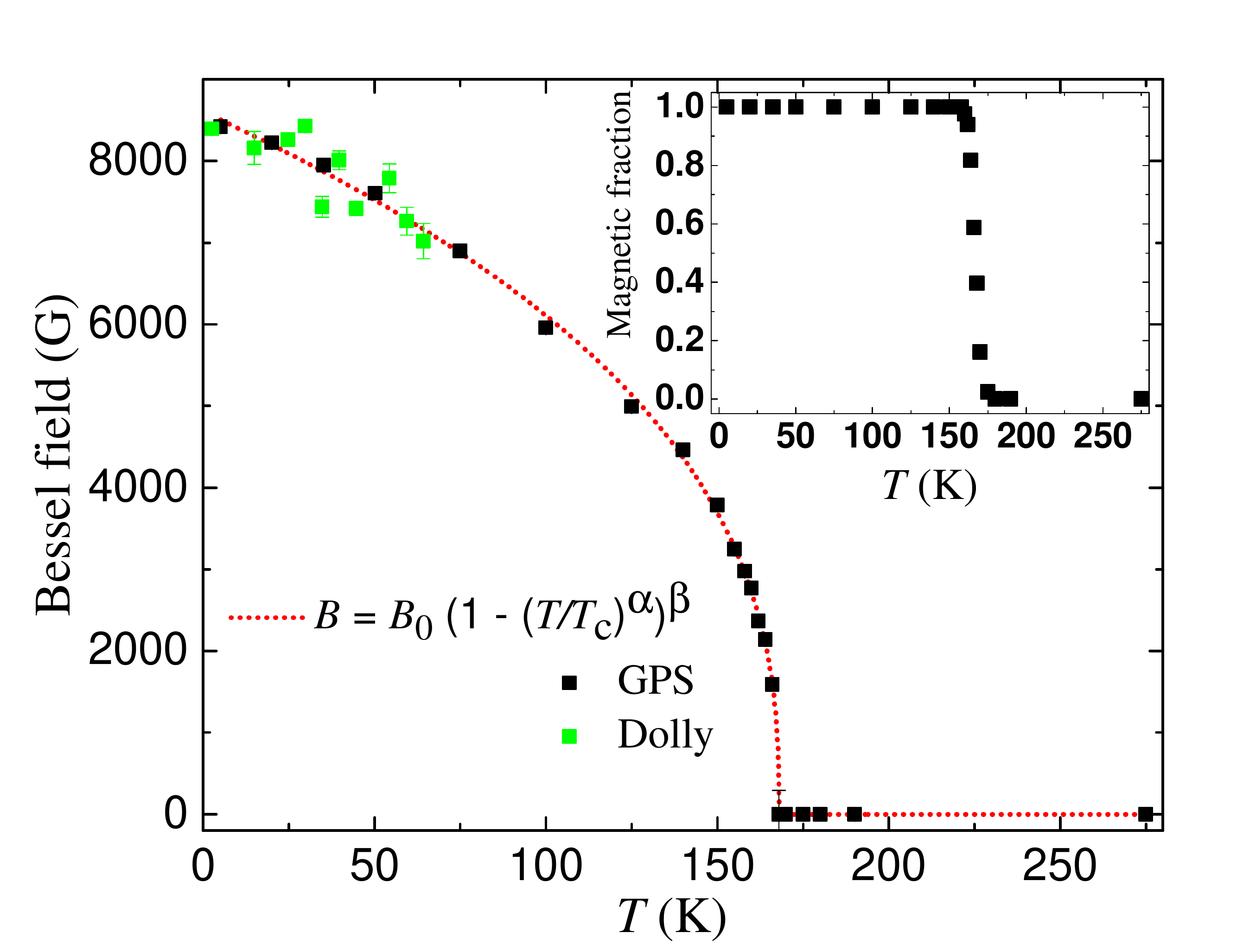}
\caption{\label{fig:orderpm_fraction} Main panel shows the temperature dependence of the Bessel field as determined from the fit. Lines indicate the phenomenological fit model as detailed in the text. Inset depicts the magnetic volume fraction as a function of temperature.}
\end{figure}

The temperature dependence of the Bessel field $B$ and the magnetic fraction $f_\mathrm{mag}$ are shown in Fig.~\ref{fig:orderpm_fraction}. The magnetic volume fraction $f_\mathrm{mag}$ increases to 100\% in a temperature range of 15\,K and is fixed to 100\% below 125\,K for simplicity. A similar behavior is observed by 30\,mT TF measurements (not shown). The Bessel field $B$, which is for a temperature independent magnetic structure proportional to the magnetic order parameter, is fitted by $B(T) = B_{0} [1-(T/T_{\mathrm{N\mu}})^{\alpha_{1}}]^{\beta_{1}}$, yielding a zero temperature value $B_{0}$ = 0.860(8)\,T, a transition temperature $T_{\mathrm{N\mu}}$ = 168\,K and a critical exponent $\beta_{1}$ = 0.378(7). The empirical parameter $\alpha_{1}$ is introduced to describe the data at temperatures much below the transition temperature, but it is not really needed in the present case as the fit yields $\alpha_{1}$ = 1.00(6). 
The experimental $\beta$ value is very close to the theoretical value of  0.362 for 3D Heisenberg magnets which is consistent with the expected behavior of the Fe$^{3+}$ moments. 

\begin{figure}[h]
\includegraphics[width=\columnwidth]{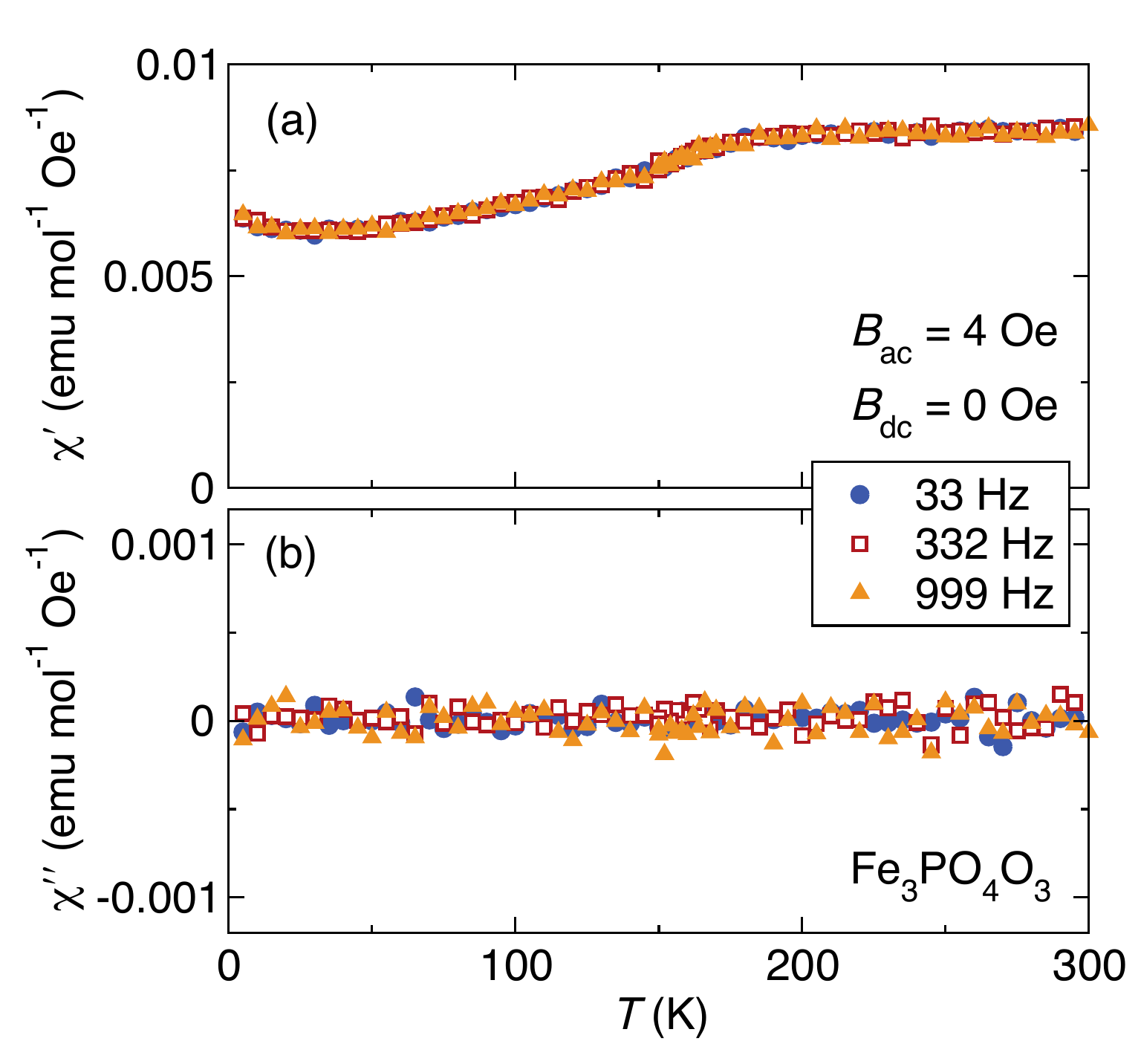}
\caption{\label{fig:ac-sus} Temperature variation of the real and imaginary part of the ac magnetic susceptibility at different frequencies.}
\end{figure}

Figure~\ref{fig:ac-sus} shows the real and imaginary parts of the ac magnetic susceptibility measured at different frequencies. The strength of the ac signal is overall very weak, and resembles the dc magnetic susceptibility.~\cite{PhysRevB.92.134419} A slight dip in $\chi^\prime$ is observed at the expected N$\mathrm{\acute{e}}$el transition, while no features are observed in the out-of-phase susceptibility ($\chi^{\prime\prime}$). No other temperature-dependent anomalies or frequency-dependent relaxation processes are observed in the ac susceptibility.

\begin{figure}[h]
\includegraphics[width=\columnwidth]{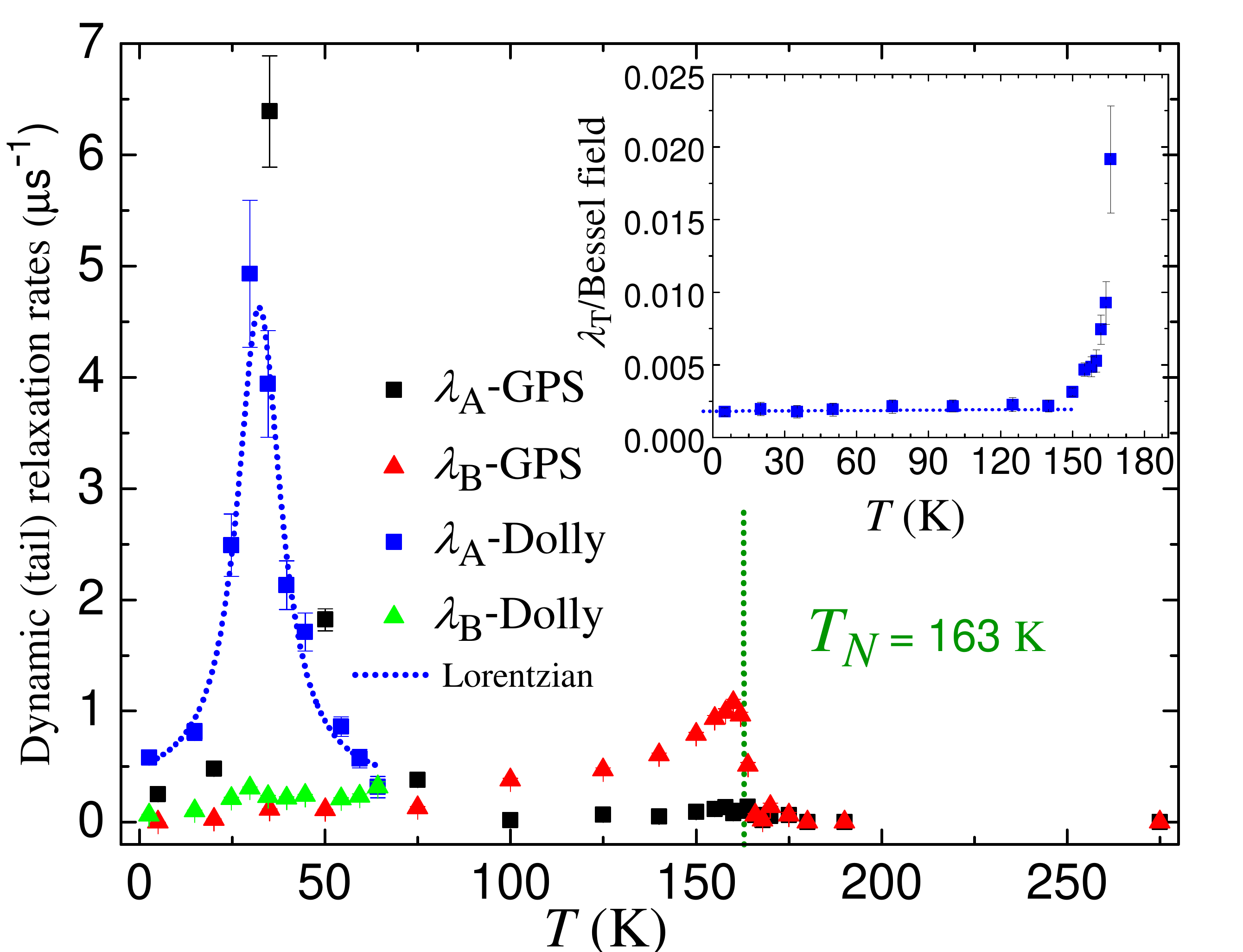}
\caption{\label{fig:relaxation} Temperature dependence of the ZF-\muSR\ longitudinal/tail relaxation rates (main panel) and of the normalized transverse relaxation rate $\lambda_\mathrm{mag}$ (inset).}
\end{figure}

So far, we have discussed mainly the static properties of the \fp\ system. In the following we focus our discussion on the dynamic properties manifested in the relaxation of the longitudinal part of the ZF-$\mu$SR spectra. As mentioned before, a full description of the tail at all temperatures requires a sum of two exponential [cf. Eq.~(\ref{eqn:muonAsymmetry})]. A global fit taking into account the spectra recorded at different temperatures simultaneously yields a weight of $f_\mathrm{A}=$ 0.33 for site A and =0.66 for site B. Figure~\ref{fig:relaxation} shows the ZF tail relaxation rates $\lambda_\mathrm{tail,A}$ and $\lambda_\mathrm{tail,B}$ as a function of temperature. Upon lowering the temperature, as the system approaches the paramagnetic to helical magnetic phase transition at $\sim$163\,K, maximum of $\lambda_\mathrm{tail,B}$ $\approx 1.07$ $\mu$s$^{-1}$ and $\lambda_\mathrm{tail,A}$ $\approx 0.1 \mu$s$^{-1}$ is observed. Surprisingly, a strong peak of $\lambda_\mathrm{tail,A}$ is observed at 35\,K, well below the magnetic transition for the site A only. There is no significant change of the ac susceptibility response at 35\,K, however. This might be attributed to the different time scales probed by the two techniques: ac susceptibility measures fluctuations on a much slower time scale (Hz) than \muSR (MHz). 
\\
The behavior of this second peak in the ZF-\muSR relaxation rate evidently represents a non critical type, as it can be well fitted by a Lorentz function. Similar features were also observed in the antiferromagnet Cobalt Glycerolate and the partially frustrated magnet a-Fe$_{92}$Zr$_{8}$.~\cite{{Pratt-PhysRevLett.99.017202},{Lierop-PhysRevLett.86.4390}} While in the former case the authors propose the motion of domains as a cause, transverse spin freezing to a noncollinear state with coexistence of ferromagnetic and spin-glass order is promoted in the later case. Although the available neutron scattering results do not have fine enough steps in the low temperature region to be conclusive there is no sign of any second phase transition in  \fp. However, it is worth mentioning that there is an increase of the relative intensity of the sharp magnetic peak near 1.35 \AA\ between 100\,K and 4\,K, which may represent the lack of full ordering until below 100\,K. Moreover, since there is no anomaly in the magnetic order parameter at 35\,K as determined both from the \muSR\ and M\"ossbauer (cf. next section) studies a sudden modification of the magnetic structure can be ruled out in the temperature regime below 100\,K. Therefore, the effect of AFM domain wall motion is the most likely explanation for the peak in $\lambda_\mathrm{tail,B}$. Below this second relaxation peak all the fluctuations freeze and the rates approach zero as $T\rightarrow$0.

In the inset of Fig.~\ref{fig:relaxation} the transverse relaxation rate $\lambda_\mathrm{mag}$ normalised by the Bessel field $B$ is plotted as a function of temperature. This is a way of measuring the homogeneity of a magnetic ordering. Below 140\,K, $\lambda_\mathrm{mag}/$(Bessel field) remains constant ruling out any further change of the sublattice magnetization. However, above 140\,K the continuous increase of the ratio $\lambda_\mathrm{mag}/$(Bessel field) towards $T_\mathrm{N}$ implies an increase in static disorder as the magnetic coherence length decreases, consistent with the neutron scattering data.
    	
\section{ $^{57}$Fe M\"ossbauer results}

\begin{figure}[h]
	\centering
		\includegraphics{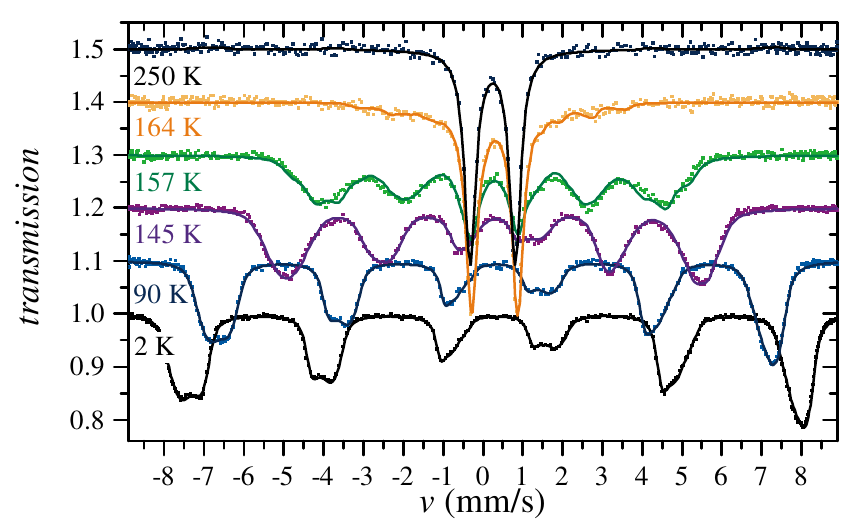}
	\caption{Representative zero-field M\"ossbauer spectra at different temperatures as indicated. The spectra show a complex line shape due to a distribution of absolute value and orientation of the magnetic hyperfine field with respect to the EFG. This causes a special shape for each individual line.}
	\label{spectra}
\end{figure}

In order to investigate the magnitude and the orientation of the ordered static Fe magnetic moment,
$^{57}$Fe M\"ossbauer spectroscopy was carried out.
Given that the $^{57}$Fe nucleus probes the on-site magnetism of Fe, any significant changes of orbital and spin degrees of freedom should be reflected in the $^{57}$Fe M\"ossbauer studies.
The M\"ossbauer technique offers an accurate estimation of the single ion magnetic order parameter, via the measured hyperfine field ($B_{hyp}$) at the Fe site. Since the $B_{hyp}$ is measured relative to the electric field gradient (EFG) subtle changes of the magnetic order parameter and/or magnetic structure can be resolved.
\\
\indent Figure~\ref{spectra} shows the zero field $^{57}$Fe M\"ossbauer spectra at representative temperatures. At high temperatures in the paramagnetic state, the \fp\ data consists of a single symmetric doublet with a quadrupole splitting of -1.13\,mm/s and a room temperature centre shift of 0.461(1)\,mm/s, which is consistent with the high spin Fe\textsuperscript{3+} state.
We have used a large absorber thickness, and we observed an absorber line width which is close to the spectrometer minimum line width. This suggests a good sample quality. Due to the half filled valence orbitals no valence contribution to electric field gradient (EFG) is expected, accordingly it arises from the lattice contribution only. The nearest neighbor Fe environment consists of five O\textsuperscript{2-} ions.

The neutron scattering results indicate the presence of a tiny amount of impurity phase Fe\textsubscript{2}O\textsubscript{3} in \fp. In our present M\"ossbauer data, a 1.2(1) \% Fe$_2$O$_3$ fraction could be resolved in the spectra with highest statistics and this is consistent with neutron diffraction~\cite{0022-3719-19-35-018}. Subsequently this tiny fraction was neglected to reduce the number of parameters for the analysis.

Below the transition temperature ($T_{50\%}=161(1)$\,K, $T_{\mathrm{NM}}=164.8(5)$\,K) the \fp\ shows a Zeeman splitting with a saturation value of 47.9\,T corresponding to 3.7\,$\mu_{\mathrm{B}}$ if pure core contribution of the Fermi contact interaction is assumed.~\cite{eriksson1989isomer,novak2010contact} However, the  M\"ossbauer spectra in the ordered state do not represent a single sextet and instead reveal a complex pattern suggesting distributions of the angle $\theta$ between the local magnetic hyperfine field $B_{\mathrm{hyp}}$ and the EFG $z$-axis and/or a distribution of $B_{\mathrm{hyp}}$ (see the 2\,K spectra in Fig.~\ref{spectra}).

\begin{figure}[h]
	\centering
		\includegraphics{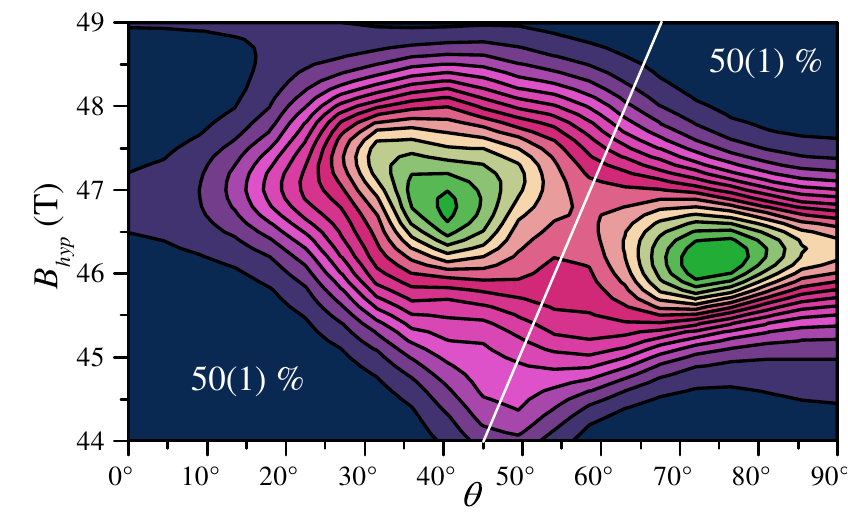}
	\caption{Maximum entropy method applied to the 30\,K spectrum to deduce the distribution of angles with respect to EFG z-axis and magnitude of local magnetic fields. Distribution of polar angle $\theta_\mathrm{Bhyp-Vzz}$ and absolute value $B_\mathrm{hyp}$ of the magnetic hyperfine field with respect to the EFG z-axis. The grid of possible $B\time\theta$ pairs consists of $51\times 21$ values. The distribution was smoothed by maximum entropy method. It shows roughly a 1:1 decomposition and a systematic decrease of $B_\mathrm{hyp}$ with increasing $\theta$.}
	\label{MEM2D}
\end{figure}

\begin{figure}[h]
	\centering
		\includegraphics{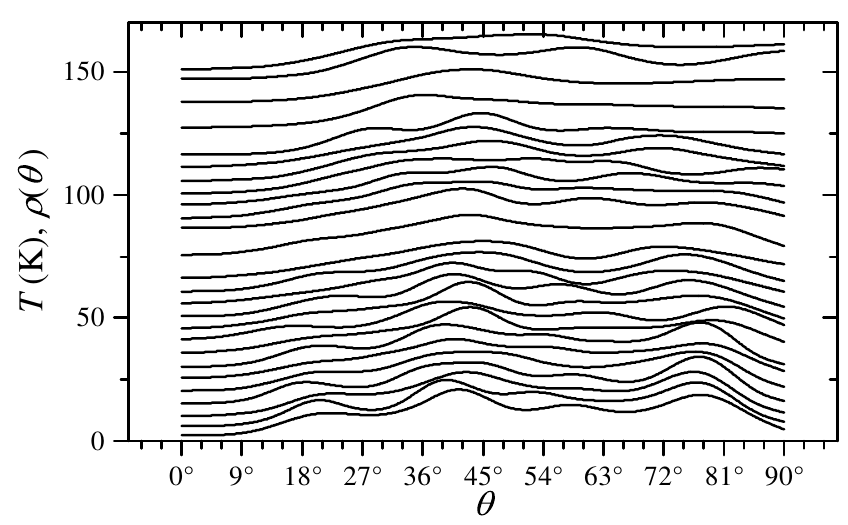}
	\caption{Distribution of the polar angle $\theta$ between the EFG-z-axis and the magnetic hyperfine field $B_\mathrm{hyp}$ with 31 equidistant sampling points. A linear dependence of $B_\mathrm{hyp}$ on $\theta$ as described in by Eq.~\eqref{theta_Bhyp_correlation} was assumed. For low temperatures and high statistics the features of the individual $\theta$ distribution become more pronounced.}
	\label{MEMtheta_individual}
\end{figure}

\begin{figure}[h]
	\centering
		\includegraphics{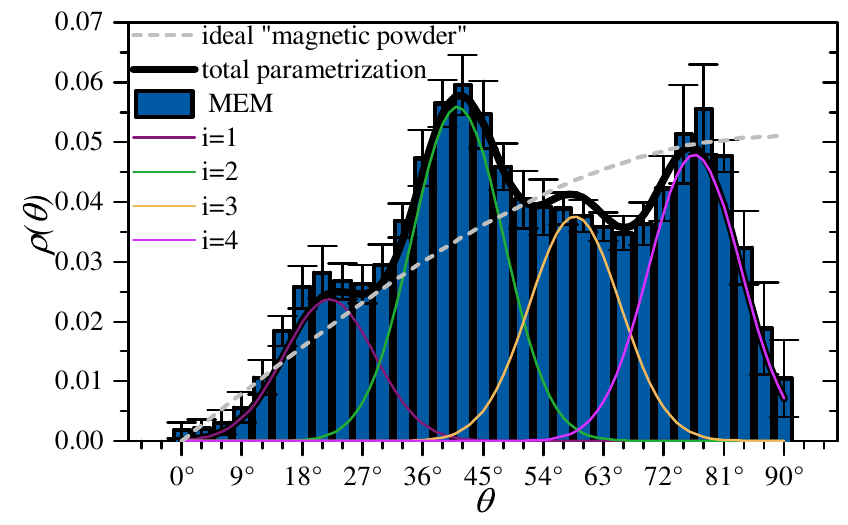}
	\caption{Average distribution of the angle $\theta$ between EFG z-axis and the magnetic hyperfine field. The distribution can be parametrized using four Gaussian peaks (parameters in table \ref{rhoparameters}). This parametrization was used to determine the magnetic order parameter $B_\mathrm{hyp}(T)$ shown in Fig.~\ref{Bhyp}.
	}
	\label{thetaMEM}
\end{figure}

In general, M\"ossbauer powder spectra are analysed by using the full static Hamiltonian:

\begin{align}
	H_s = &\frac{eQ_{\mathrm{zz}}V_{\mathrm{zz}}}{4I(2I-1)}\left[(3I_z^2-I^2)+\frac{\eta}{2}(I_+^2+I_-^2)\right] \notag\\
				&-g_I\mu_NB_{\mathrm{hyp}}\left(\frac{I_+e^{-i\phi}+I_-e^{+i\phi}}{2}\sin\theta+I_z\cos\theta\right)
				\label{eqn:Moss-pwder-hamiltonian}
\end{align}

with nuclear spin operators $I_z$, $I_{+}\,=\,I_x+iI_y$, and $I_{-}\,=\,I_x-iI_y$, where ${B_{hyp}}$ is the hyperfine field at the $^{57}$Fe site. $Q$, $g_\mathrm{I}$, and $\mu_\mathrm{N}$ indicate the nuclear quadrupole moment, g factor and magneton, respectively. The polar angle $\theta$ and the azimuthal angle $\phi$ represent the orientation of the Fe hyperfine field $B_\mathrm{hyp}$ with respect to the EFG $z$ axis.
As it can be seen from the Eq.~\eqref{eqn:Moss-pwder-hamiltonian}, in powder M\"ossbauer spectroscopy the static Hamiltonian depends only on $V_{\mathrm{zz}}$, $B_{\mathrm{hyp}}$ and $\theta$ if an axial symmetric EFG is assumed. Because $V_{\mathrm{zz}}$ turned out to be well defined in the paramagnetic states, a probability distribution $\rho(\theta,B)$ (Fig.~\ref{MEM2D}) was deduced by means of the maximum entropy method (MEM) assuming equal isomer shift and EFG at every Fe nucleus. This distribution reveals the existence of two different Fe environments (green colored areas in Fig.~\ref{MEM2D}) and both parameters appear to be distributed to some extent. Moreover, $B_{\mathrm{hyp}}$ and $\theta$ seem to be anti-correlated. The fit model was reduced to a single $\theta$ distribution, assuming 
\begin{align}
B_\mathrm{hyp}(\theta)=B_0(1+\epsilon\theta), 
\label{theta_Bhyp_correlation}
\end{align}

where $\epsilon$ is -0.297 mT/$^{\circ}$. The magnetic spectra were fitted by means of $\theta$-MEM (Fig.~\ref{MEMtheta_individual}). The average distribution $\rho(\theta)$ is shown in Fig.~\ref{thetaMEM}. The characteristic 4-peak structure becomes less pronounced with increasing temperature.

Since MEM cannot be combined with a transmission integral fit we parametrized $\rho(\theta)\propto\sum_i f_ie^{-((\theta-\theta_i)/\sigma)^2/2}$ by four Gaussian distributions with equal width $\sigma$. The parameters $f_i$ and $\theta_i$ (table \ref{rhoparameters}) were obtained from a global fit of all temperatures below 110\,K. This fit results in an angular distribution width $\sigma=6.8^\circ$.

\begin{table}[h]
\centering
\begin{tabular}{lcccc}\\\hline
i&1&2&3&4\\\hline
$f_i$&0.145&0.337&0.220&0.297\\
$\theta_i$&21.8$^\circ$&41.0$^\circ$&58.6$^\circ$&76.2$^\circ$\\\hline
\end{tabular}
\caption{Parameters of the distribution of the angle $\theta$ between $B_{\mathrm{hyp}}$ and EFG-z-axis $\rho(\theta)\propto\sum_i f_ie^{-(\theta-\theta_i)/\sigma)^2/2}$.}
\label{rhoparameters}
\end{table}

The Gaussian parametrization was applied to every single run, and it fits the whole temperature range (Fig.~\ref{thetaMEM}). For temperatures close to the magnetic transition it was necessary to allow a standard deviation $\sigma_{\mathrm{hyp}}$ of $B_{\mathrm{hyp}}$. Figure~\ref{Bhyp} shows the magnetic order parameter as determined from the above analysis, and can be adequately described by a two exponent model, where the classical critical exponent is $\beta_2=0.24$.
The transition temperature is 165\,K determined from the order parameter fit.

We have also performed field dependent M\"ossbauer experiments (see appendix~\ref{sec:LF-Moessbauer}) to study the effect of magnetic moments to the field. But no such strong field dependency is observed. 

\section{Discussion}

Consistent with previous measurements we observe a high spin Fe\textsuperscript{3+} state that orders magnetically below $164\pm1$\,K.
Given that the EFG is almost axially symmetric, only a distribution of $\theta$ and $B_{\mathrm{hyp}}$ can account for the complex M\"ossbauer spectrum. It was shown that the total distribution can be projected to a pure theta distribution assuming a weak linear dependence $B_\mathrm{hyp}(\theta)$ due to a small orbital dependence. Such a theta distribution can be sufficiently parametrized by a superposition of four Gaussian peaks.

To derive a microscopic magnetic structure from the $\theta$ distribution in the hexagonal lattice it is important to consider the orientation of the EFG-$z$ with respect to the crystallographic $c$-axis. For further discussion we introduce a spherical coordinate system with the polar angle $\zeta$ and the aximuthal angle $\alpha$, where $\zeta=0$ specifies the crystallographic $c$-axis and $\zeta=90^\circ, \alpha=0^\circ$ an axis in the $ab$-plane pointing from P to Fe in topview. According to a point charge calculation the EFG-$z$-axis is given by $\zeta=-37^\circ$ and $\alpha= 0^\circ$ as shown in Fig.~\ref{LatteiceEFG} (see appendix~\ref{sec:EFG_Calculation}).

The N$\mathrm{\acute{e}}$el type skyrmion and a simple one-dimensional-correlation model, both on the basis of neutron diffraction data, were tested for consistency with the M\"ossbauer data, i.e. the extracted $\rho(\theta)$ distribution (see appendix~\ref{sec:thetaDistribution}).
Both magnetic structures could roughly reproduce $\rho$, however a direct fit of these models with pitch fit to the M\"ossbauer data is not satisfying. As a final step, we assumed a ferromagnetic alignment of the three Fe moments in a triangular \fp\ unit and extracted the distribution of the triangle's moment direction (see appendix~\ref{sec:MeM} for details). These investigations suggest that the magnetic moments tend to point parallel along the $c$-axis, instead of showing in the $a$-$b$-direction. A helical plane including the $c$-axis seems improbable because it requires an equal distribution of the polar angle $\zeta$, which is not supported by the data analysis. The discrepancy of the temperature dependence of the local fields measured by $\mu$SR and M\"ossbauer spectroscopy can qualitatively be understood by a less pronounced orientation of the moments towards the $c$-axis for higher temperatures. However if the previous simplifications made were believed to be true, then one may consider a continuous rotation by the angle $\alpha$ of the tilted moments around the $c$-axis. In that case the distribution of $\theta(\zeta_0,\alpha)$ with respect to the tilting angle $\zeta_0$ of the EFG $z$-axis and the opening angle $\zeta$ with respect to the $c$-axis is given by the following expression.

\begin{align}
\cos\theta(\alpha,\zeta,\zeta_0)&=\sin\zeta_0\cos\alpha\sin\zeta+\cos\zeta_0\cos\zeta\\
\rho(\theta,\zeta_0,\zeta)&\propto \frac{\partial}{\partial \theta} \alpha(\theta)\\
&= \frac{\sin\theta}{\sqrt{\sin^2\zeta\sin^2\zeta_0-(\cos\theta-\cos\zeta\cos\zeta_0)^2}}
\label{label_of_eq_5}
\end{align}

Such a distribution has two pronounced (diverging) peaks corresponding to the experimentally observed pairs of angles.

\begin{figure}[ht]
	\centering
		\includegraphics{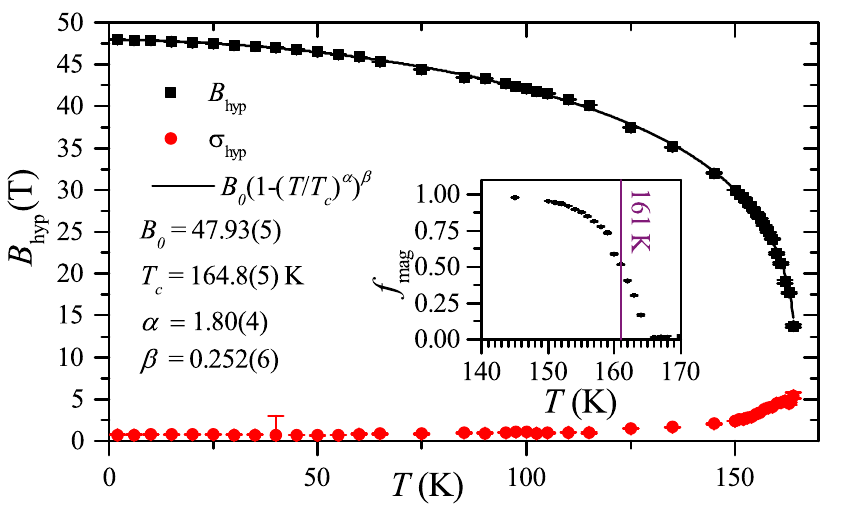}
	\caption{Temperature dependence of the magnetic hyperfine field $B_{\mathrm{hyp}}$ and its standard deviation $\sigma_\mathrm{hyp}$. Inset shows the magnetic volume fraction as determined by M\"ossbauer experiments.}
	\label{Bhyp}
\end{figure}

Below 120\,K, the static order parameter obtained from the \muSR\ and M\"ossbauer data appear to be inconsistent, while just below $T_N$ they seem to agree with each other (cf. Fig.~\ref{fig:orderpm_fraction} and \ref{Bhyp}). The M\"ossbauer hyperfine field (order parameter) is more direct measurement of the Fe single ion order parameter as M\"ossbauer probes the onsite magnetism of $^{57}$Fe, while in \muSR\ experiments the muons experience transferred hyperfine fields. It is therefore more indirect. 
Consequently, a change in the helix pitch angle without a significant change of the magnetic moment can explain the more pronounced temperature dependence of the order parameter seen by \muSR\ at low temperatures compared to the M\"ossbauer data.

\section{Conclusions}

In conclusion we have performed detailed \muSR\ and M\"ossbauer studies on the \fp\ powder system. We did observe a homogeneous long range magnetic ordering with 100 \% magnetic volume fraction. Both \muSR\ and M\"ossbauer results are in line with the helical type of ordered magnetic structure as already proposed by Ross \textit{et al.} Ref.~[\onlinecite{PhysRevB.92.134419}]. In addition, we investigated the dynamic properties of \fp. From a \muSR\ point of view we found a second spin-lattice relaxation peak at around 35\,K (0.21 $T_{\mathrm{N}\mu}$), apart from the peak at $T_{\mathrm{N}\mu}$. This second peak might be associated with the low energy scales of \fp\ or it can also be the effect of domain wall motion. However, for having a better understanding of this system single crystal studies are needed. 
The M\"ossbauer spectra are modeled by taking into account the specific angular distribution of the local hyperfine field $B_{\mathrm{hyp}}$ with respect to the local electric field gradient. Further, the N$\mathrm{\acute{e}}$el type skyrmion and a simple one-dimensional-correlation model were examined for consonance with M\"ossbauer data with the extracted $\rho(\theta)$ distribution.

\subsection{ACKNOWLEDGMENTS} This research is partially supported by the Deutsche Forschungsgemeinschaft (DFG) through SFB 1143 for the project C02 and through the priority program SPP 1458 (KL 1086/10-2). S.H. gratefully acknowledges the financial support by the Swiss National Science Foundation (SNF-Grant No. 200021-159736). V.G. is thankful to DFG for the financial assistance through the GR 4667/1-1.

\appendix

\section{\label{sec:EFG_Calculation}Calculation of the electric field gradient}

\begin{figure}[ht]
	\centering
		\includegraphics[width=\columnwidth]{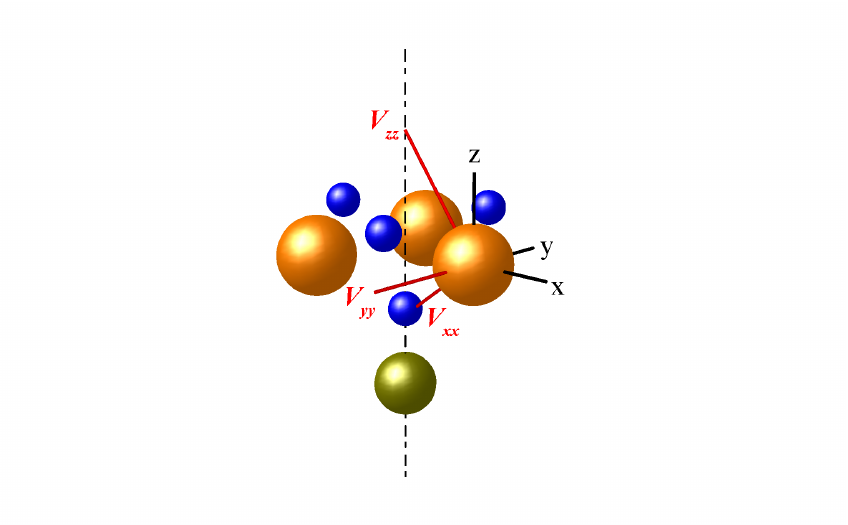}
	\caption{Triangular Fe\textsubscript{3}PO\textsubscript{4} unit with phosphorous (green), oxygen (blue) and  iron (orange). The Euclidean coordinate system centered at an iron atom is shown in black. The principal coordinate system of the EFG $V_{xx}-V_{yy}-V_{zz}$ at the Fe position with the angle $\zeta_0=-37^\circ$ between EFG-z-axis and normal z-axis is shown in red.
}
	\label{LatteiceEFG}
\end{figure}

In this report the angles $\zeta$ and $\alpha$ are polar and azimuthal angles, where $\zeta=0$ specifies the crystallographic c-axis and $\zeta=90^\circ,\alpha=0^\circ$ an axis in the ab-plane pointing from P to Fe in topview. A point in the Euclidian coordinate system is given by $(x,y,z)=r(\cos\alpha\sin\zeta,\sin\alpha\sin\zeta,\cos\zeta)$ (Fig.~\ref{LatteiceEFG}).

\begin{figure}[ht]
	\centering
		\includegraphics[width=\columnwidth]{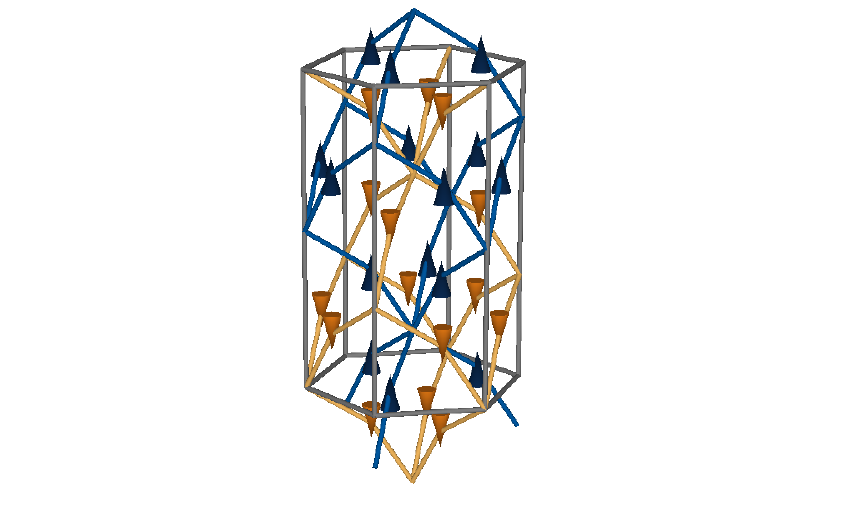}
	\caption{Magnetic unit cell of the parent AFM sublattices. Fe moments are represented by cones, sticks connect P and Fe. 
	}
	\label{MagneticUnitCell}
\end{figure}

The EFG was recalculated using Moessfit's \textit{CrystalFields::Fe3PO4O3\_Vii}. It assumes point charges Fe\textsuperscript{3+}, P\textsuperscript{5+} and O\textsuperscript{2-} on the positions specified in Ref.~\onlinecite{PhysRevB.92.134419}. Because Fe\textsuperscript{3+} is in high spin state, this lattice contribution should be sufficient. $N=\pm 50$ unit cells in each direction were considered, however less then 1\,\% deviation is achieved for $N >\pm 14$. The following principal axes are found $V_{zz}=-2.63$\,V/\AA\textsuperscript{2} at $\zeta_0=-37^\circ$ with a small $\eta=0.18$. The EFG-z-axis thus points to the center of a triangular Fe\textsubscript{3}PO\textsubscript{4} unit. In the same way the EFG x-axis does. This means that the EFG-y-axis lies within crystallographic ab-plan. The situation is sketched in Fig.~\ref{LatteiceEFG}.

The calculated $V_{zz}$ value is 62\,\% smaller than the measured value, assuming a Sternheimer antishielding factor \cite{chen2007mossbauer} of $\gamma_\infty=-9.14$ for \textsuperscript{57}Fe\textsuperscript{3+}.

\section{\label{sec:thetaDistribution}$\rho(\theta)$ distribution for different micro-magnetic structures}

 The earlier (not shown here) deduced $\rho_\mathrm{MEM}(\theta)$ distribution of the angle $\theta$ between EFG-$z$-axis and $B_\mathrm{hyp}$ can be interpreted in terms of a conical modulation along the $c$-axis with distinct opening angles of $40^\circ$ and $59^\circ$ using Eq. \eqref{label_of_eq_5}. However, such modulation is based on $\zeta_0\approx \pm 18^\circ$ which is not reasonable. However, the pronounced peak of $\rho$ at $\theta=42^\circ$ allows for a different interpretation: The system may tend to a commensurate parent structure with moments parallel to $c$, i.e. $\theta\approx\zeta_0$. Little structural deviations from the diffraction data could allow for such correspondence. Even more important is the distance of $\Delta\theta=35^\circ$, which is close to the tilting which is expected from the pitch of $|\delta|=0.073$\,\AA$^{-1}$.~\cite{PhysRevB.92.134419} This pitch leads to a tilting of $33.7^\circ$.

\begin{figure}[ht]
	\centering
		\includegraphics{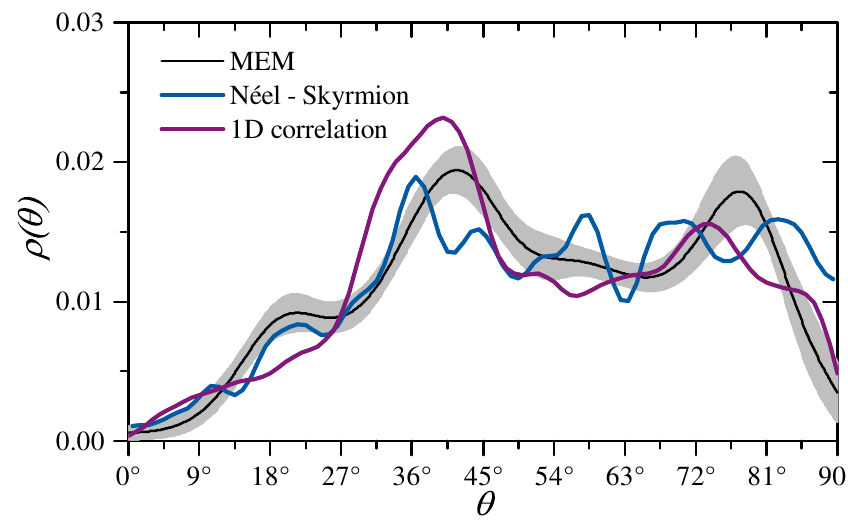}
	\caption{Comparison of the angular distribution of distinct micro-magnetic structures and the experimentally determined $\rho(\theta)$ from Fig.~\ref{thetaMEM}. The position of the largest peak and the angular difference to the second largest peak can be roughly understood assuming parallel orientation of the moments to the crystallographic $c$-axis and the pitch of 0.073\,\AA\textsuperscript{-1} as proposed by neutron diffraction measurements, irrespective of the radial weight function. 
	}
	\label{CompareDistributions}
\end{figure}

This consideration suggests a N$\mathrm{\acute{e}}$el-type skyrmion with the axis along $c$, centered at a P-position. A coherence length of $\xi=70$\,\AA{} and a hexagonal supercell with a diameter of 20 unit cells was assumed. The fraction of properly rotated moments is given by $e^{-d/\xi}$ with the distance $d$ from the center. It gives a peak at $\theta=d\delta_0$. The incoherent fraction $(1-e^{-d/\xi})$ leads to a powder like distribution $\propto\sin\theta$. The resulting distribution is shown in Fig.~\ref{CompareDistributions}. $\rho_\mathrm{MEM}$ can be reproduced much more accurately by small variation of $\xi$ and $\delta_0$, however the best fit is very sensitive to tiny changes of the parameters, thus this model was discarded.

A more simple approach to reproduce $\rho_0$ is a 1D-correlation. Similar to the Skyrmion the angle $\theta=d\delta_0$ depends on the distance $d$ from the starting moment which is considered to point in c-direction. The contribution of the angles are weighted by $e^{-d/\xi}$, and the behaviour of incoherent moments is not considered. Such model reproduces two pronounced peaks in the $\rho(\theta)$ distribution (Fig.~\ref{CompareDistributions}).

Both models were implemented in Moessfit to directly fit the M\"ossbauer spectra with the parameters $\delta_0, \xi$ and $\zeta_0$. The goodness of fit is not convincing and moreover is highly sensitive to tiny changes of the parameters.

\section{\label{sec:MeM}$\zeta-\alpha$ maximum entropy method (MEM)}

\begin{figure}[ht]
	\centering
		\includegraphics{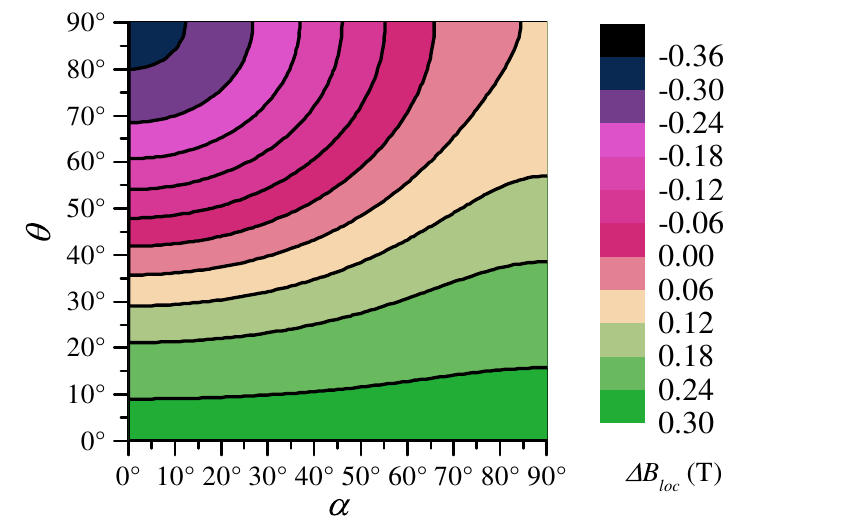}
	\caption{
	Variation of the local magnetic field due to dipole fields of the two neighboring Fe moments in a triangular
Fe$_3$PO$_4$ unit with the polar angle $\theta$ and the azimuthal angle $\alpha$ as explained at the beginning of appendix A. Eventually, the systematic $\theta$-dependence of $B_\mathrm{hyp}$ can be explained by such dipole fields. These dipole fields are essential in the $\xi-\alpha-$MEM described in appendix C.}
		\label{dBloc}
\end{figure}

In both approaches used in appendix~\ref{sec:thetaDistribution} the intensity of peaks was basically a result of a fixed starting direction parallel to the c-axis and the coherence $e^{-d/\xi}$. However, another intrinsic structural reason could be responsible for that: the $120^\circ$ rotational symmetry. According to neutron diffraction a helical $120^\circ$ rotation is ruled out. Instead, a parallel alignment of the moments within one triangular Fe\textsubscript{3}PO\textsubscript{4} can be assumed and described by $\zeta$ and $\alpha$. For such a group of three iron moments the dipole fields can be calculated. It turns out, that the local field can vary by $\pm 0.3$\,T due to a pure dipole contribution (Fig.~\ref{dBloc}), assuming 4.2 $\mu_B$/Fe.

Such a variation was observed in the $B_\mathrm{hyp}-V_{zz}$-MEM (Fig.~\ref{MEM2D}) and used in the fit model for $\rho_0(\theta)$. A maximum entropy method (MEM) was applied to the reduced $\zeta\times\alpha=[0^\circ,90^\circ]\times[0^\circ,60^\circ]$-space, which still contains full information due to symmetry reasons. For each $(\zeta,\alpha)$ pair three subspectra are generated. For each sub-spectrum different $\theta$ values and the local magnetic fields are calculated. To meet the required field variation the value of the moment was declared as a fit parameter, the fits lead to $m\approx 10 \mu_B$ which is more than twice the actual value. This may be explained by additional transferred hyperfine fields of the neighbouring irons, which act on top of the dipole fields. The sub-spectra are weighted with $\sin\zeta$ according to the area element of the spherical coordinates. The \textit{SHpLev} theory function of Moessfit was used to level the sub-spectra according to saturation effect. Still this is only an approximation because transmission integrals and MEM cannot be used simultaneously, but it provides a large improvement compared to a thin absorber model. Asymmetry of the EFG is neglected ($\eta=0$).

\begin{figure}[ht]
	\centering
		\includegraphics{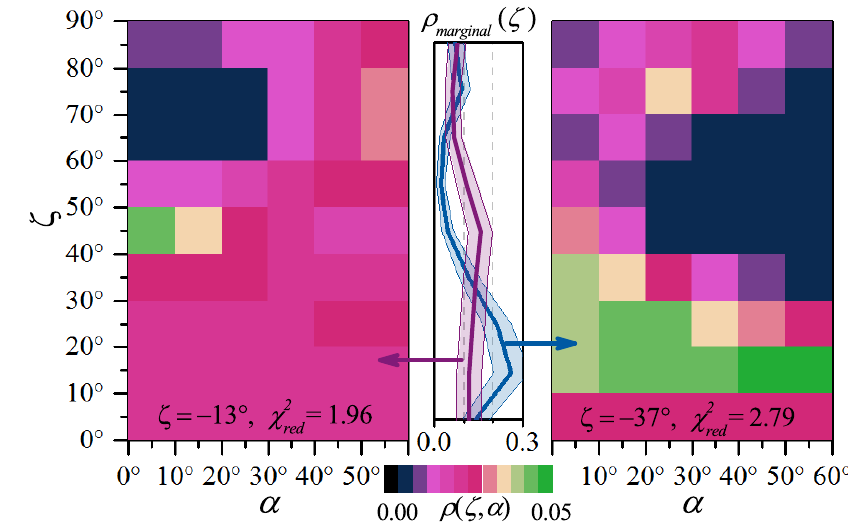}
	\caption{Orientation of the moment of a triangular Fe\textsubscript{3}PO\textsubscript{4} unit at $T=30$\,K described in spherical coordinates $(\zeta,\alpha)$. $\rho(\zeta,\alpha)$ is obtained by MEM. Two scenarios are presented: one with fixed tilting angle $\zeta_0=-37^\circ$ and another where this angle is a fit parameter. The errors are confined to $\Delta\rho(\zeta_i,\alpha_i)<0.025$. The marginal probability $\rho_\mathrm{marginal}$ (eq.~\eqref{rhoamrgianl}) is incompatible with an equal distribution.}
	\label{combined_small-N}
\end{figure}

The $\zeta\times\alpha$-space is discretized in $9\times6$ sampling points, so that MEM is applied to $54=9\cdot 6$ sub-spectra. The result is shown for $T=30$\,K in Fig.~\ref{combined_small-N}.

\begin{figure}[ht]
	\centering
		\includegraphics{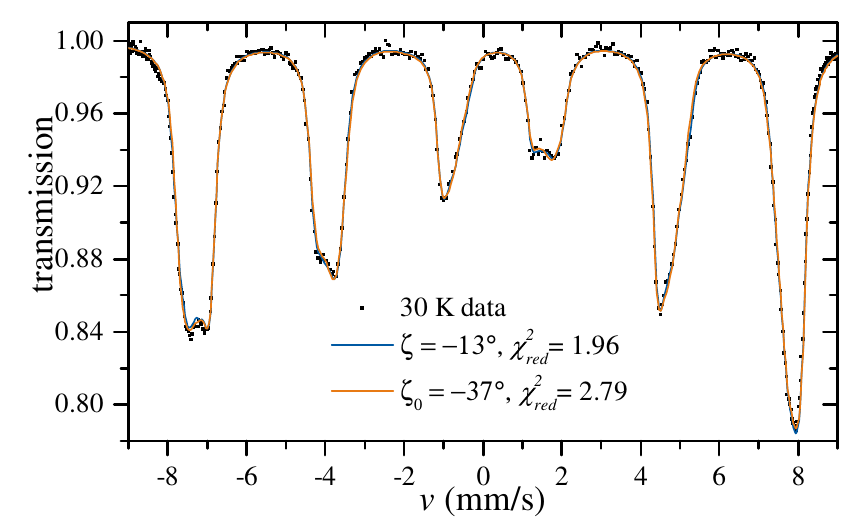}
	\caption{30\,K data fitted by the $\zeta-\alpha$-MEM with two different tilting angles $\zeta$ of the principal axis of the EFG.}
	\label{small-N-spectra}
\end{figure}

Two different cases are shown: for one $\zeta_0=-37^\circ$ was fixed, for the other it is a fit parameter, leading to $\zeta_0=-13^\circ$. This additional degree of freedom significantly improves the fit, but both models yield satisfying agreement with the data (Fig. \ref{small-N-spectra}). Aside the better fitting the $V_{zz}=-69$\,V/\AA\textsuperscript{2} leads to a smoother temperature dependence with regard to paramagnetic $V_{zz}$ values than $V_{zz}=-73$\,V/\AA\textsuperscript{2} for the fixed $\zeta=-37^\circ$. In Fig. \ref{combined_small-N} the marginal probability
\begin{equation}
	\rho_\mathrm{marginal}(\zeta)=\sum_{i=1}^6 \rho(\zeta,\alpha_i)
	\label{rhoamrgianl}
\end{equation}

is shown. It is in any case incompatible with an equal distribution of $\zeta$ as it would result from a helical plane including the c-axis, which was suggested by Ross \textit{et al.}
in Ref.~\onlinecite{PhysRevB.92.134419}. Even though the $\zeta_0=-13^\circ$ restriction is close to a random distribution of axes of moments, it still has a preferred and an avoided direction. For the restricted $\zeta_0=-37^\circ$ case the density $\rho$ is focused around $\zeta=18^\circ$ (better visible if 486 sampling points are used) with only little $\alpha$-dependence, which in principle is consistent with a conical axis in c-direction and an opening angle of $18^\circ$.

\begin{figure}[ht]
	\centering
		\includegraphics[width=\columnwidth]{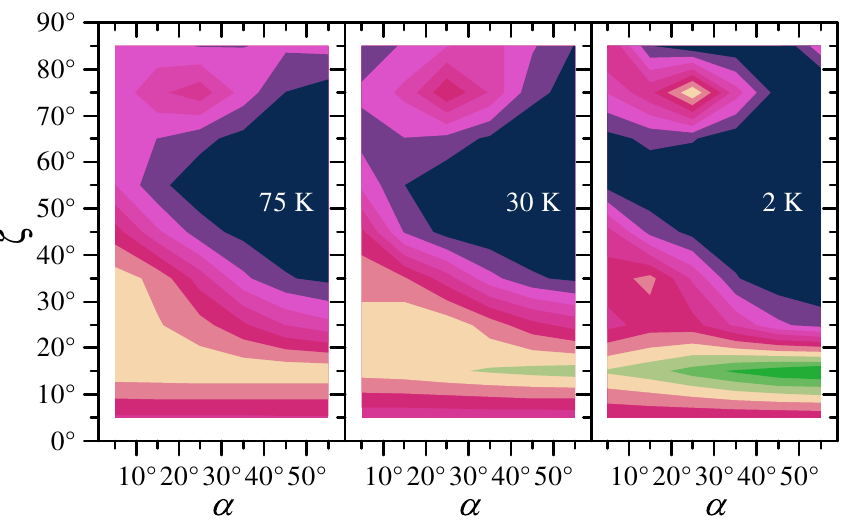}
	\caption{Temperature dependence of $\rho(\zeta,\alpha)$ with $\zeta_0=-37^\circ$ similar to Fig. \ref{combined_small-N} with a larger number of $\zeta-\alpha-$sampling points. }
	\label{T-dependence}
\end{figure}

The distribution $\rho(\zeta,\alpha)$ is shown for different temperatures in Fig. \ref{T-dependence}. It in principle supports the picture of a transition to a more commensurate spin structure, or alternatively to larger coherence length (which is incompatible with neutron diffraction).

\section{\label{sec:dipole-field}Muon site estimation on \fp\ using dipole field calculation}

\begin{figure}[ht]
	\centering
		\includegraphics[width=\columnwidth]{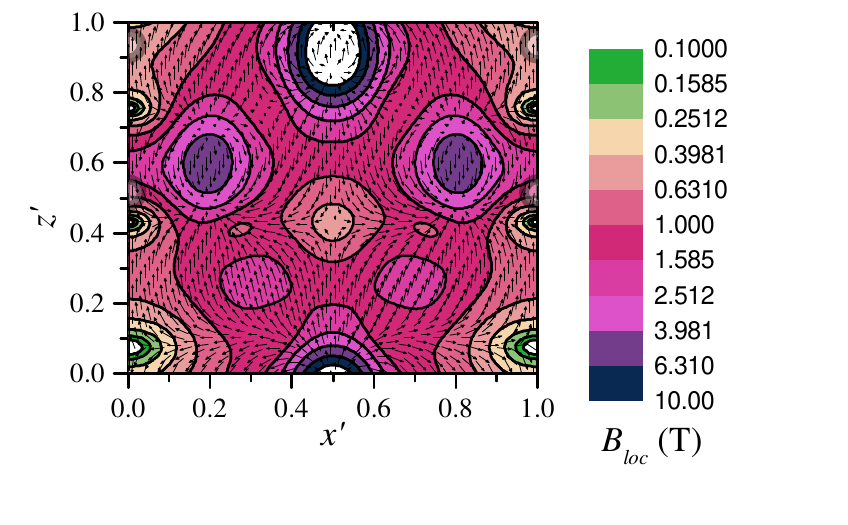}
	\caption{Dipole field of the proposed parent magnetic structure, plane: $\vec{r}=x'\vec{a}+z'\vec{c}$}
	\label{x-z-cut}
\end{figure}

\begin{figure}[ht]
	\centering
		\includegraphics[width=\columnwidth]{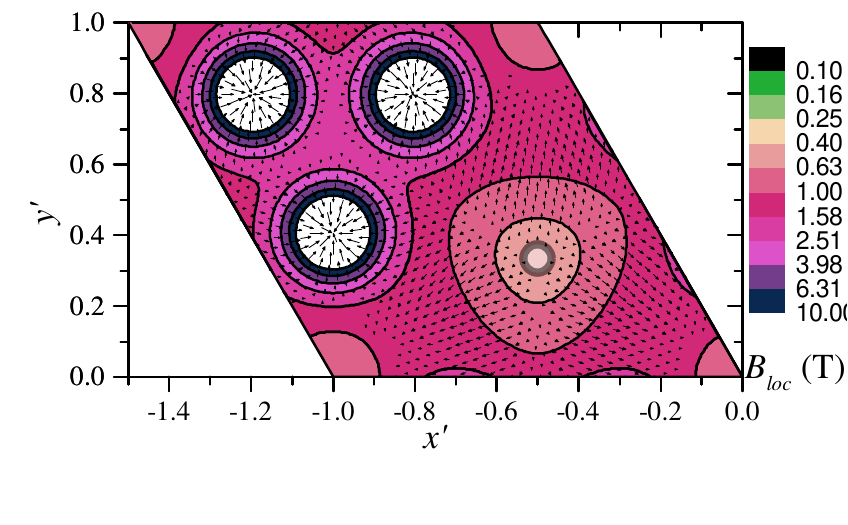}
	\caption{Dipol field of the parent magnetic structure,  plane: $\vec{r}=x'\vec{a}+y'(\vec{b}+\vec{a}/2)+(1/3-0.05)\vec{c}$}
	\label{x-y-cut}
\end{figure}

\begin{figure}[ht]
	\centering
		\includegraphics[width=\columnwidth]{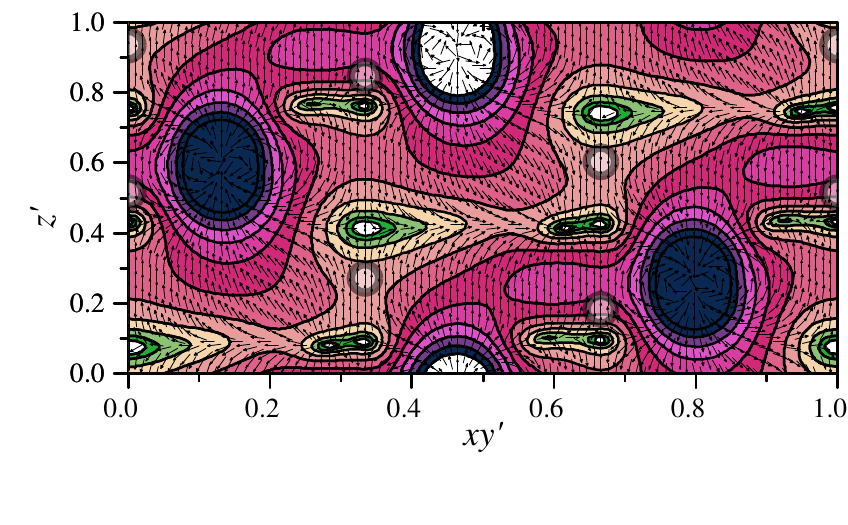}
	\caption{Dipol field of parent magnetic structure, plane: $\vec{r}=\vec{a}+xy'(\vec{b}-\vec{a})+z'\vec{c}$}
	\label{xy-z-cut}
\end{figure}

Dipole field calculations were performed assuming magnetic moments of 4\,$\mu_\mathrm{B}$ at the iron site. Considering the parent AFM structure as depicted in Fig.~\ref{MagneticUnitCell}, the estimated local fields are shown in the cuts through the crystallographic unit cell in Fig.~\ref{x-z-cut}, \ref{x-y-cut}, and \ref{xy-z-cut}. For $x^\prime$\,=\,$y^\prime$\,=\,0 there are P atoms at $z^\prime$\,=0 and $z^\prime$\,=1. The muon is supposed to reside on the high symmetry axis connecting these P (or alternatively at one of the O\textsuperscript{2-} nearby). The local field of 0.8\,T sensed by the muon allows for two muon sites, which are sketched by circles in the figures. However, the one muon position close to P is very unlikely due to the positive charge of P\textsuperscript{5+}.


\begin{figure}[ht]
	\centering
		\includegraphics[width=\columnwidth]{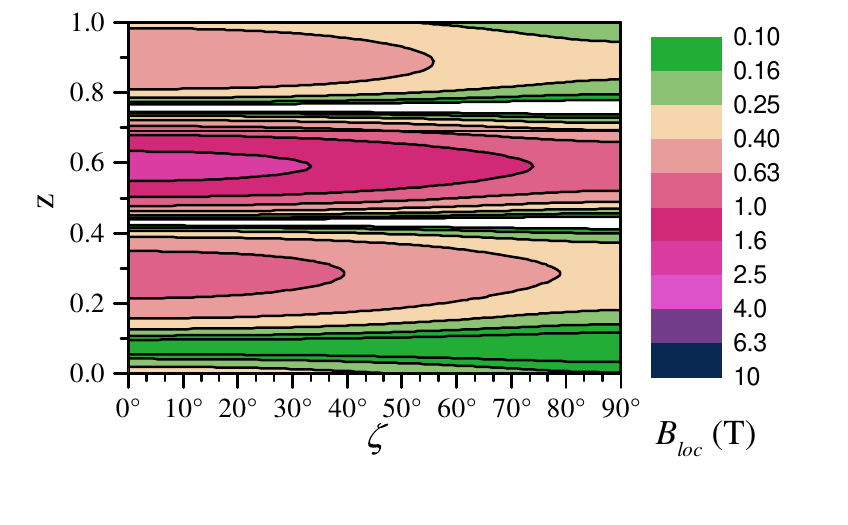}
	\caption{Dependence of the local magnetic field along $z'\vec{c}$ i.e. along the axis of possible muon position on tilting by the angle $\zeta$ of the magnetic moments.}
	\label{MBS-Modell}
\end{figure}

The variation of the local field is studied as a function of polar tilting angle $\zeta$ of the moments with respect to the c-axis, assuming the same orientation of all up moments, and the same orientation of all down-moments, respectively, is shown in Fig. \ref{MBS-Modell}. Due to the $120^\circ$ rotational symmetry of the axis at $x$\,=\,$y$\,=\,0, there is no azimuthal dependence of the absolute value $|\vec{B}_\mathrm{loc}|$ of the local field. The one important piece of information of Fig. \ref{MBS-Modell} is the decreasing value of $B_\mathrm{loc}$ with increasing tilting $\zeta$, which in principle could explain the flatter temperature dependence of the $\mu$SR magnetic order parameter compared to M\"ossbauer spectroscopy.

\begin{figure}[ht]
	\centering
		\includegraphics{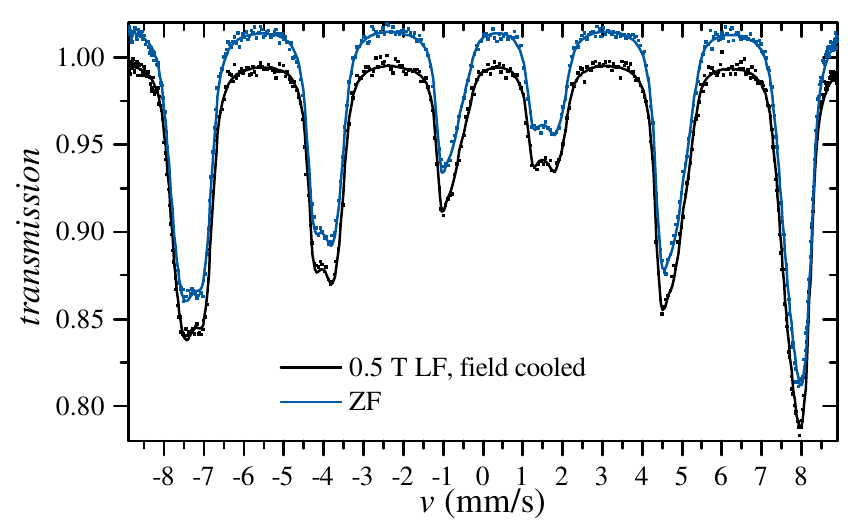}
	\caption{M\"ossbauer spectra at 25\,K at zero and 0.5\,T applied field show no difference.}
	\label{LFspectra}
\end{figure}

\section{\label{sec:LF-Moessbauer} Low-Field M\"ossbauer experiments}

A small field of 0.5\,T was applied longitudinal to the gamma beam and the sample was field-cooled. This field causes only small source splitting and the measurement geometry has not to be changed. The intention of the LF experiment was to study the reaction of the magnetic moments to the field and eventually observe a field induced transformation of the magnetic structure. However as shown in Fig. \ref{LFspectra} no such changes are observed. Even the LF spectra are well described by the above described zero-field model with the only difference of a small increase of $B_\mathrm{hyp}$ by 0.2\,\%.

\bibliography{Fe3PO4O3}

\end{document}